\documentclass[9pt,twocolumn,twoside]{opticajnl}
\journal{opticajournal} 

\setboolean{shortarticle}{true}

\usepackage{upgreek}
\usepackage{lineno}

\title{A tunable FP Perot coupling interferometer on thin-film lithium niobate}

\author[1,*]{Ayed Al Sayem}
\author[1]{Heqing Huang}
\author[1]{Ting-Chen Hu}
\author[1]{Alaric Tate}
\author[1]{Mark Cappuzzo}
\author[1]{Rose Kopf}
\author[1]{Mark Earnshaw}

\affil[1]{Nokia Bell Labs, NJ, United States}

\begin{abstract}
We experimentally show an electro-optic tunable Fabry-Perot cavity on thin-film lithium niobate (TFLN). Instead of tuning the cavity phase and thus resonant frequency, we demonstrate modulation of the FP cavity by tuning the cavity mirrors via the electro-optic effect in the couplers. We enable full tuning of a low-Q FP cavity with only 3.5\,V Vpi and a very short 3.5\,mm long Mach-Zehnder interferometer (MZI) mirror.
\end{abstract}

\setboolean{displaycopyright}{false} 

\begin{document}

\maketitle

\section{Introduction}

Thin-film lithium niobate (TFLN) and tantalte (TFLT) are currently one of the most promising material platforms for both classical and quantum photonic technologies \cite{TFLN_review_Zhu2021TFLNReview}. Both TFLN and TFLT platforms have been used for high-speed electro-optic (EO) modulation \cite{Lwang2018integrated,TFLT_kWang2023_LiTaO3PIC} and non-linear optical parametric processes \cite{Lwang2018ultrahigh,Lzhang2019broadband}. Resonant devices can enhance the modulation \cite{xue2022breaking} and non-linear conversion efficiencies \cite{Lu2021,kundu2024periodically}. Resonators, especially micro-ring resonators, have been extensively explored on the TFLN platform \cite{Lzhang2017monolithic} for numerous applications such as frequency comb generation \cite{Lzhang2019broadband}, optical parametric oscillator \cite{Lu2021}, electro-optic modulation \cite{xue2022breaking}, and so on. In general, high-Q cavities are desired for many applications as they provide strong interaction between the interacting modes \cite{Lu2021,mckenna2022ultra}. However, they also make the devices sensitive to environmental fluctuations, as well as material-induced instability \cite{xu2021mitigating}. Fabry-Perot (FP) micro-cavities are less explored on the TFLN platform \cite{Zhang2023SingleMode,Guo2023EOTunable}, mainly due to the slightly larger size and lower quality factors compared to the micro-ring cavities \cite{Zhang2023SingleMode,Guo2023EOTunable}. However, for many practical applications, such as wavelength selection for hybrid lasers \cite{Lfranken2025high}, electro-optic modulation \cite{li2020lithium}, and erbium-doped lasers \cite{Zhang2023SingleMode}, one does not require an ultra-high Q; instead, stability is more important. FP cavities also operate in transmission mode, which can be desirable for applications such as frequency comb generation and on-chip lasers. In this paper, we propose and experimentally demonstrate a thermo-optic (TO) and electro-optic (EO) tunable Fabry-Perot (FP) cavity based on a cascaded Mach-Zehnder interferometer (MZI) with loop mirrors. The reflectivity of the mirrors of the FP cavity is controlled by the phase difference of the MZI interferometer, which is controlled by either fast EO or slow but stable TO tuning. We show both TO and EO coupling modulation of the FP cavity. The design is fabrication-tolerant, as the MZI acts as a tunable splitter for each reflector. Finally, we show efficient modulation of the cavity transmission by EO modulation with an effective Vpi of 3.5\,V using only 3.5\,mm long MZI modulator with widely spaced electrodes.     

\begin{figure*} [!ht] 
    \centering
    \includegraphics[width=0.70\textwidth]{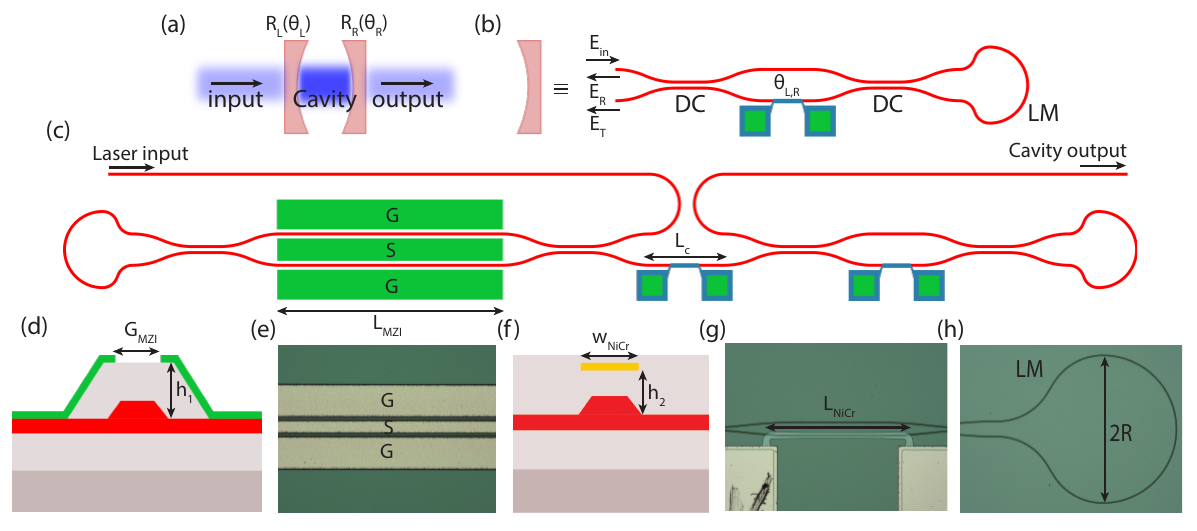}
    \caption{(a) Basic FP cavity constructed by two reflectors with reflectivity $\mathrm{R}_{\theta_{L}}$ and $\mathrm{R}_{\theta_{R}}$. (b) Equivalent photonic integrated circuit (PIC) diagram for one of the reflector. (c) Schematic of the PIC of the FP cavity. (d), (f) Cross-section of the electro-optic (EO) and thermo-optic (TO) section of the FP cavity, here the parameters are $\mathrm{h_{1}}=1.5\,\mu\mathrm{m}$, $\mathrm{G_{MZI}}=6.5\,\mu\mathrm{m}$. Optical microscope image of the (e) CPW electrodes, (g) NiCr heater on top of the MZI waveguide, and (h) photonic loop mirror (LM). Here, $\mathrm{w_{NiCr}}=10\,\mu\mathrm{m}$, $\mathrm{h_{2}}=1\,\mu\mathrm{m}$, $\mathrm{L_{NiCr}}=130\,\mu\mathrm{m}$, and $\mathrm{R}=100\,\mu\mathrm{m}$.  DC: directional coupler, LM: loop mirror, G: ground, S: signal.}  
    \label{Fig1}
\end{figure*}

\section{Device design}
Fig.\,\ref{Fig1}(a) shows a basic FP cavity constructed by reflective mirrors with reflectivity $\mathrm{R}_{\theta_{L}}$ and $\mathrm{R}_{\theta_{R}}$. Fig.\,\ref{Fig1}(b) shows the equivalent photonic integrated circuit (PIC) diagram for one of the reflectors. The reflector can be constructed by connecting the output ports of the MZI via a loop mirror. The MZI is constructed by connecting two identical directional couplers (DCs) with a length of $\mathrm{L_{DC}}$, and gap $\mathrm{g_{DC}}$ and waveguide section with a length, $\mathrm{L_{MZI}}$. The reflectivity of the mirror can be tuned by changing the phase difference, $\delta_{\theta_{L,R}}$, between the MZI waveguides either by thermo-optic (TO) or electro-optic (EO) tuning. Fig.\,\ref{Fig1}(c) shows the schematic image of the proposed device. The device comprises two reflectors (left and right) connected by a waveguide with a length, $\mathrm{L_{C}}$, which acts as the cavity. The right reflector consists of an MZI with TO control, where resistive heaters are placed on top of one of the arms of the MZI waveguide to thermally tune the phase by changing the refractive index of the waveguide, hence the reflectivity of the mirror. The left reflector consists of an MZI with EO control, where the CPW electrodes with a length of $\mathrm{L_{MZI}}$ is used to tune the phase of the MZI in a push-pull configuration \cite{Lwang2018integrated}. Fig.\,\ref{Fig1}(d) and Fig.\ref{Fig1}(f) show the cross-section of the EO and the TO tuning section, respectively. The signal and ground electrodes are placed on top of the LN waveguides with a gap, $\mathrm{G_{MZI}}$, as shown in Fig.\,\ref{Fig1}(d). For the thermal tuning, a NiCr-based resistive heater with a length of $\mathrm{L_{NiCr}}$ is placed on top of the LN waveguide with an oxide clad between the LN waveguide and the resistive heater. Fig.\,\ref{Fig1}(e),(g), and (h) show the optical microscope image of the GSG electrodes, the MZI section with TO tuners, and the loop mirror, respectively. Two bus waveguides are used as the input and output waveguide couplers for the FP cavity.

\section{Device Fabrication}
The fabrication process starts with a 4-inch 600\,nm thick TFLN wafer on thermally grown $4.7\,\mu \mathrm{m}$ thick oxide on a silicon substrate commercially available from Nano-LN. Photonic device components such as Waveguides, directional couplers (DCs), and loop mirrors are photolithography defined and subsequently etched using $\mathrm{Ar^{+}}$ plasma dry etch. The waveguides are partially etched up to $\sim$325\,nm, leaving a slab of $\sim$275\,nm. Using plasma-enhanced chemical vapor deposition (PECVD), $1\,\mu\mathrm{m}$ thick oxide layer is then deposited as cladding. NiCr heaters with sheet resistivity, $\rho=\sim18\,\Omega/\square$, and $1\,\mu\mathrm{m}$ thick co-planar waveguide gold (Au) electrode layers are then defined by photolithography and subsequent etching or lift-off process. NiCr heaters are passivated by another $0.5\,\mu\mathrm{m}$ thick oxide. The wafer is diced, and the individual chips are polished for efficient fiber-to-chip coupling.  

\section{Measurement results}
\begin{figure}[!ht]
  \centering
  \includegraphics[width=9cm]{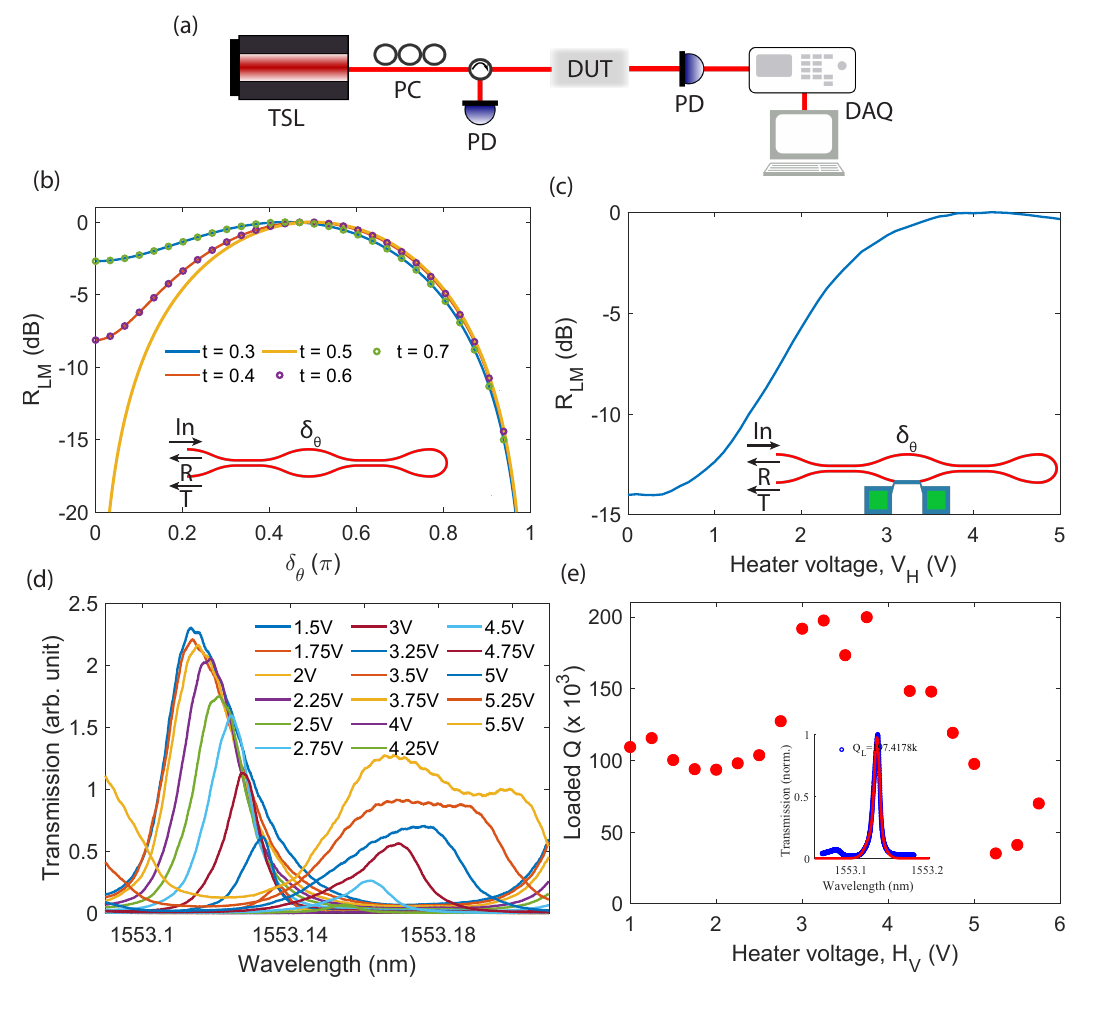}
\caption{(a) Schematic of the measurement setup. TSL: tunable scanning laser, PC: polarization controller, PD: photo-detector, DUT: device under test, DAQ: data acquisition. (b) Numerically calculated reflection of the loop-mirror as a function of the phase difference of the MZI arms for different values of the DC strength, $\mathrm{t_{DC}}$. (c) Experimental values of the reflection of the LM as a function of the heater voltage. (d) Transmission of the FP cavity as a function of wavelength with different heater bias voltages applied to both mirrors of the FP cavity. (e) Fitted loaded Q, i.e., quality factor of the FP cavity as a function of heater voltage. The inset shows one of the resonances with a Lorentzian fit.}
\label{Fig2}
\end{figure}

\begin{figure*}[!ht]
  \centering
  \includegraphics[width=12cm]{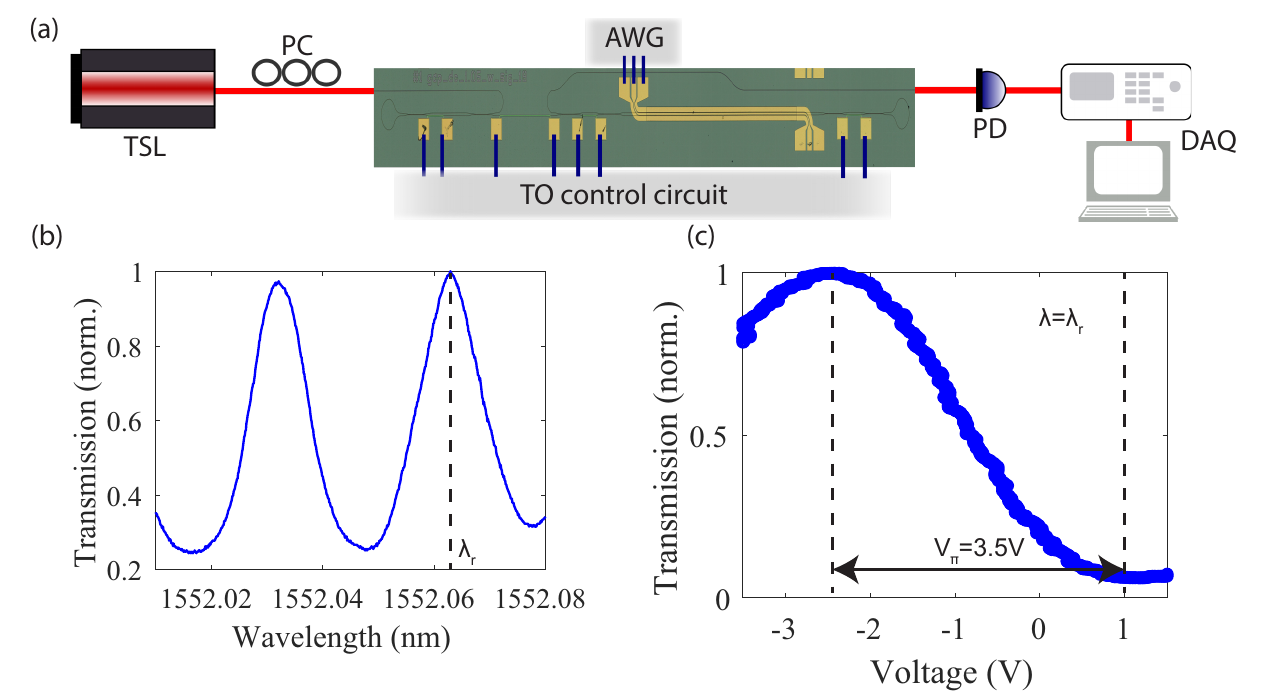}
\caption{(a) Schematic diagram of the measurement setup. PC: polarization controller, AWG: arbitrary wave-function generator, PD: photo-detector. (b) Normalized transmission of the EO tunable FP cavity as a function of wavelength. The mirror reflectivity is set such that the resonance is slightly over-coupled. (c) Normalized transmission of the cavity resonance as a function of the applied voltage (V). Here, the laser set is at ${\lambda=\lambda_{r}}$.}
\label{Fig3}
\end{figure*}

First, we characterize the reflectivity of one of the mirrors of the FP cavity using just thermal tuners, where the reflectivity of both the left and right mirrors is tuned by the TO phase shifters instead of the EO phase shifters. Fig.\,\ref{Fig3}(a) shows the schematic measurement setup. Light from a tunable laser (Santec-570) is coupled to the DUT using a lensed fiber through a fiber-optic polarization controller (PC) and a fiber-optic circulator (Thorlabs). The output light from the DUT is collected by another lensed fiber and sent to a photo-detector (PD). The reflected light from the DUT is collected by the input lensed fiber and is sent to another PD through the reflective port of the circulator. We first measure the reflected light from just one of the mirrors of the FP cavity as a function of the bias of the MZI. The bias or the phase difference of the MZI arms is controlled by the on-chip resistive thermal heaters. In Fig.\,\ref{Fig2}(b), we plot the theoretical reflectivity of a single loop-mirror (LM) as a function of the phase difference between the MZI arms, with different values of the coupling strength of the directional coupler (DC), $\mathrm{t_{DC}}$. 

For an ideal DC, one would expect $\mathrm{t_{DC}}$ to be \,50\%. Unfortunately, because of the fabrication imperfections such as the variation of the film thickness, etch depth, and the waveguide geometry, the coupling strength of the DC, $\mathrm{t_{DC}}$, can vary from the target value of \,50\% \cite{huang2023fabrication}. However, using the MZI as a tunable splitter, one can achieve \,50\% splitting ratio even with an imperfect DC as long as $\mathrm{t_{DC}}$ falls within \,15\% to \,85\% \cite{Miller2015Perfect,Miller2017Meshes}.  In Fig.\,\ref{Fig2}(c), we plot the reflection from the LM, $\mathrm{R_{LM}}$ as a function of the bias voltage of the heater. From the zero bias condition, we can estimate $\mathrm{t_{DC}}$ to be $\mathrm{\simeq0.55}$. In Fig.\,\ref{Fig2}(d), we show the transmission of the FP cavity as a function of wavelength with different heater bias voltages applied to both heaters for the left and the right mirror. This device only has thermal tuners, and we applied the same voltage to both heaters. Fig.\,\ref{Fig2}(e), we plot the loaded quality factor, Q of the resonances after fitting with a Lorentzian curve. As expected, with the increase of the heater voltage, the mirrors become more reflective and hence, the loaded Q increases and peaks at 
bias voltage where the mirror reflectivity is maximized as can be observed from Fig.\,\ref{Fig2}(c). The inset of Fig.\,\ref{Fig2}(e) shows the resonance at the maximum reflectivity condition. The loaded Q of the resonances can be tuned from $\mathrm{\sim\,40k}$ to $\mathrm{\sim\,200k}$ which corresponds to 3\,dB bandwidth tuning from $\mathrm{\sim\,4.8\,GHz}$ and $\mathrm{\sim\,0.97\,GHz}$.  
Fig.\,\ref{Fig3}(a) shows the schematic of the measurement setup for the EO tuning of the FP cavity. The measurement setup is similar to the one shown in Fig.\,\ref{Fig2}(a). Here, the reflectivity of the left mirror is controlled by the TO phase shifters and the reflectivity of the right mirror is controlled by both TO and EO phase shifters. An RF signal generated from an AWG at 1\,MHz is applied to the CPW electrodes of the MZI of the right reflector. The cavity is biased at a slightly over-coupled condition. Fig.\,\ref{Fig3}(b) shows the transmission of the FP cavity as a function of wavelength. The laser wavelength is then set to one of the resonant wavelengths at $\lambda=\lambda_{r}$ as shown in Fig.\,\ref{Fig2}(b). Fig.\,\ref{Fig3}(c) shows the normalized transmission of the FP cavity as a function of applied voltage.
Complete modulation can be achieved with just 3.5\,V with a 3.5\,$\mu$m long modulation section with a gap, $\mathrm{G_{MZI}=6.5\,}\mu$m. The theoretical VpiL product with the same geometry is $\sim$3.8\,Vcm.
Modulation efficiency can be drastically increased by placing electrodes close to the LN waveguide \cite{Lwang2018integrated,Lzhang2021integrated}, which will reduce VpiL product and hence the Vpi of the FP cavity without increasing the modulation length. 
During the preparation of this manuscript, a recent study by Qi et al.\cite{Qi:25} reported related findings with a similar design proposed in this letter, but with only thermal modulation.

\section{Conclusion}
In conclusion, we demonstrate an electro-optic (EO) and thermo-optic (TO) tunable FP cavity on TFLN. Such devices can have fundamental applications, including high-speed EO modulation, on-chip fast-tuning lasers, directly modulated lasers, wavelength division multiplexing, non-linear frequency conversion, and quantum light generation.   

\begin{backmatter}
\bmsection{Funding} 
Nokia Corporation of America.

\bmsection{Disclosures} 
Nokia Corporation of America. Patent application, 333399-US-NP.

\bmsection{Data availability} Data underlying the results presented in this paper are not publicly available at this time but may be obtained from the authors upon reasonable request.

\bigskip

\end{backmatter}

\bibliography{sample}

\begin{thebibliography}{10}
\newcommand{\enquote}[1]{``#1''}

\bibitem{TFLN_review_Zhu2021TFLNReview}
D.~Zhu, L.~Shao, M.~Yu, \emph{et~al.}, {\protect\JournalTitle{APL Photonics}} \textbf{6}, 030901 (2021).

\bibitem{Lwang2018integrated}
C.~Wang, M.~Zhang, X.~Chen, \emph{et~al.}, {\protect\JournalTitle{Nature}} \textbf{562}, 101 (2018).

\bibitem{TFLT_kWang2023_LiTaO3PIC}
C.~Wang, Z.~Li, J.~Riemensberger, \emph{et~al.}, {\protect\JournalTitle{Nature}} \textbf{629}, 784 (2024).

\bibitem{Lwang2018ultrahigh}
C.~Wang, C.~Langrock, A.~Marandi, \emph{et~al.}, {\protect\JournalTitle{Optica}} \textbf{5}, 1438 (2018).

\bibitem{Lzhang2019broadband}
M.~Zhang, B.~Buscaino, C.~Wang, \emph{et~al.}, {\protect\JournalTitle{Nature}} \textbf{568}, 373 (2019).

\bibitem{xue2022breaking}
Y.~Xue, R.~Gan, K.~Chen, \emph{et~al.}, {\protect\JournalTitle{Optica}} \textbf{9}, 1131 (2022).

\bibitem{Lu2021}
J.~Lu \emph{et~al.}, {\protect\JournalTitle{Optica}} \textbf{8}, 539 (2021).

\bibitem{kundu2024periodically}
M.~Kundu, B.~Sikder, H.~Huang, \emph{et~al.}, {\protect\JournalTitle{arXiv preprint arXiv:2408.03550}}  (2024).

\bibitem{Lzhang2017monolithic}
M.~Zhang, C.~Wang, R.~Cheng, \emph{et~al.}, {\protect\JournalTitle{Optica}} \textbf{4}, 1536 (2017).

\bibitem{mckenna2022ultra}
T.~P. McKenna, H.~S. Stokowski, V.~Ansari, \emph{et~al.}, {\protect\JournalTitle{Nature Communications}} \textbf{13}, 4532 (2022).

\bibitem{xu2021mitigating}
Y.~Xu, M.~Shen, J.~Lu, \emph{et~al.}, {\protect\JournalTitle{Optics Express}} \textbf{29}, 5497 (2021).

\bibitem{Zhang2023SingleMode}
Y.~Zhang, G.-Q. Lo, D.~T.~H. Tan \emph{et~al.}, {\protect\JournalTitle{Optics Letters}} \textbf{48}, 2660 (2023).

\bibitem{Guo2023EOTunable}
J.~Guo, W.~Jin, Y.~Zhang \emph{et~al.}, {\protect\JournalTitle{Optical Materials Express}} \textbf{13}, 2644 (2023).

\bibitem{Lfranken2025high}
C.~A. Franken, R.~Cheng, K.~Powell, \emph{et~al.}, {\protect\JournalTitle{APL Photonics}} \textbf{10} (2025).

\bibitem{li2020lithium}
M.~Li, J.~Ling, Y.~He, \emph{et~al.}, {\protect\JournalTitle{Nature communications}} \textbf{11}, 4123 (2020).

\bibitem{huang2023fabrication}
P.~Huang, K.~Chen, and L.~Liu, {\protect\JournalTitle{Optics Letters}} \textbf{48}, 1264 (2023).

\bibitem{Miller2015Perfect}
D.~A.~B. Miller, {\protect\JournalTitle{Optica}} \textbf{2}, 747 (2015).

\bibitem{Miller2017Meshes}
D.~A.~B. Miller, {\protect\JournalTitle{Optics Express}} \textbf{25}, 29233 (2017).

\bibitem{Lzhang2021integrated}
M.~Zhang, C.~Wang, P.~Kharel, \emph{et~al.}, {\protect\JournalTitle{Optica}} \textbf{8}, 652 (2021).

\bibitem{Qi:25}
L.~Qi, A.~Khalatpour, J.~F. Herrmann, \emph{et~al.}, {\protect\JournalTitle{Opt. Lett.}} \textbf{50}, 5173 (2025).

\end{thebibliography}

\end{document}